\documentclass[doublecol]{epl2}

\usepackage{amsmath}
\usepackage{bm}
\usepackage[dvips]{hyperref}
\usepackage[utf8]{inputenc}

\def\vr{{\bf r}}

\def\vF{{\bf F}}

\def\tsim{{t_{\textrm{sim}}}}

\def\ur{\utilde{r}}

\def\ur{{\bf r}}

\def\eca{\epsilon_{CA}}
\def\ecb{\epsilon_{CB}}
\def\esim{\epsilon_\textrm{sim}}

\title{Self-propulsion through symmetry breaking}

\author{Pierre de Buyl\inst{1,2} \and Alexander S. Mikhailov\inst{3} \and Raymond Kapral\inst{2}}
\shortauthor{P. de Buyl \etal}
\institute{
\inst{1}{Chemical Physics Theory Group, Department of Chemistry, University of Toronto, Toronto, Ontario, M5S 3H6 Canada}\\
\inst{2}{Center for Nonlinear Phenomena and Complex Systems, Universit{\'e} libre de Bruxelles - CP231, Av. F.D. Roosevelt 50, 1050 Brussels, Belgium}\\
\inst{3}{Abteilung Physikalische Chemie, Fritz-Haber-Institut der Max-Planck-Gesellschaft, Faradayweg 4-6, 14195 Berlin, Germany}\\
}


\abstract{%
In addition to self-propulsion by phoretic mechanisms that arises from an asymmetric distribution of reactive species around a catalytic motor, spherical particles with a uniform distribution of catalytic activity may also propel themselves under suitable conditions. Reactive fluctuation-induced asymmetry can give rise to transient concentration gradients which may persist under certain conditions, giving rise to a bifurcation to self-propulsion. The nature of this phenomenon is analyzed in detail, and particle-level simulations are carried out to demonstrate its existence.
}

\begin{document}
\maketitle

Synthetic chemically-powered self-propelled micron and nano-scale motors are interesting because of their potential applications and the fundamental challenges they present as small objects that operate in the far-from-equilibrium domain in the presence of strong fluctuations. In addition to synthetic motors that move by using chemical energy to drive non-reciprocal conformational changes, similar to many biological motors, synthetic motors without moving parts that utilize asymmetric chemical reactivity to produce motion have been constructed and studied (for reviews, see Refs.~\cite{kapral_perspective_jcp_2013,hong_et_al_pccp_2010,mirkovic_et_al_small_2010,sanchez_pumera_chem_asian_2009,wang_acsnano_2009}). Such motors operate by phoretic mechanisms where the gradient of the concentration (or other) field due to asymmetric catalysis gives rise to a force that couples to fluid flow and leads to directed motion~\cite{anderson_pof_1983,golestanian_et_al_prl_2005,juelicher_prost_epje_2009}.

Even a spherical particle with uniform catalytic activity on its surface can propel itself as a result of a symmetry-breaking bifurcation if the right conditions are met~\cite{mikhailov:97,mikhailov_calenbuhr_from_cells_to_societies_2006}. A qualitative understanding of the origin of the phenomenon can be obtained from the following considerations. Suppose the catalytic reaction $A \to B$ occurs on the sphere and further suppose both the reactant $A$ and  product $B$ molecules interact with the sphere through repulsive potentials but the $B$ potential is more strongly repulsive than that of $A$~\footnote{A similar argument can be given for attractive intermolecular interactions.}. Local concentration fluctuations can produce a transient asymmetry; however, diffusion will tend to restore symmetry on a time scale $t_D \sim R^2/D$, where $R$ is the radius and $D$ is the diffusion coefficient of the reactive molecules. Suppose a reactive fluctuation occurs locally that increases the concentration of product species $B$ and decreases the concentration of reactant species $A$ near a portion of the surface of the catalytic sphere. Since $B$ particles interact with the sphere with a stronger repulsive potential than $A$ particles, the sphere will experience a net force directed away from the area of local high $B$ concentration. (Note that since total momentum is conserved, this force is balanced by a corresponding force on the fluid.) If the catalytic sphere moves with velocity $V$ as a result of this effect, it will travel a distance $R$ in a time $t_V \sim R/V$.
When $t_V \ll t_D$ there will be insufficient time for diffusion to homogenize the concentration field around the sphere, the concentration inhomogeneity will persist and an instability can occur that gives rise to directed motion. The velocity $V$ depends on the reaction rate, particle size, and the strength and range of the interactions of the reactive species with the sphere. As these parameters vary there should be a critical condition beyond which self-propelled motion is observed. By contrast, if the product $B$ particles interact less strongly with the catalytic sphere than the $A$ reactant particles, the motion of the sphere induced by the fluctuation will be directed towards the local high $B$ concentration area and this will assist homogenization of the concentration field around the sphere by diffusion. In this case conditions are not favorable for the onset of the instability.

Self-propulsion of beads driven by actin polymerization has been observed and has some general features in common with the phenomenon described above. Spherical polystyrene beads uniformly coated with a protein that catalyzes actin polymerization can undergo a spontaneous symmetry-breaking process to induce directed motion~\cite{cameron:99}. For small beads spontaneous fluctuations are sufficient to induce symmetry breaking while larger beads move only if surface asymmetry is intentionally introduced. The symmetry-breaking mechanism has been investigated in experiments~\cite{cameron:99,reymann:11} and models~\cite{oudenaarden:99,mogilner:03,zhu:12} and depends on the detailed nature of the polymerization process. The spherical catalyst in our study is propelled by a diffusiophoretic mechanism and the symmetry-breaking mechanism differs from the actin propulsion mechanism.

The aims of this Letter are to demonstrate the existence of self-propulsion of spherical catalytic particles by a symmetry-breaking mechanism through particle-based simulations of the dynamics, to quantitatively characterize its properties, and to provide a theoretical description of the origin of the effect~\cite{michelin13}.
To this end, we consider a mesoscopic model of a chemically-active spherical particle $C$ in a fluid comprising $A$ and $B$ particles. The $A$ and $B$ molecules interact with the $C$ sphere through repulsive central Lennard-Jones potentials,
\begin{equation}
V_{C\alpha}(r) = 4 \epsilon_{C\alpha} ( \left( \frac{\sigma}{r}\right)^{12} - \left( \frac{\sigma}{r}\right)^{6} + \frac{1}{4} )\theta(r_c-r),
\end{equation}
where $\alpha \in \{A, B\}$, $\theta(r)$ is the Heaviside function and $r_c = 2^{1/6} \sigma$. The energy and distance parameters are $\epsilon_{CA}$, $\epsilon_{CB}$ and $\sigma$, respectively. In addition to these direct interactions, the $C$ sphere catalyzes the reaction $A +C \to B +C$. To implement this reaction an interaction zone around the catalytic sphere is identified and every $A$ particle that enters the interaction region is triggered for reaction. Specifically, after an $A$ particle encounters the catalytic sphere by passing through the interaction zone, its identity is changed from $A$ to $B$ as it leaves this zone. By carrying out the reaction in this way there are no changes in intermolecular potentials.

The solvent particles interact among themselves through multiparticle collision (MPC) dynamics~\cite{malevanets_kapral_mpcd_1999,kapral_adv_chem_phys_2008,gompper_et_al_adv_polym_sci_2008} where particles stream and undergo effective collisions at discrete time intervals $\tau$. The multiparticle collisions are carried out by dividing the system into a grid of cells and assigning rotation operators $\hat{\omega}_{\xi}$, chosen from a set of rotation
operators, to each cell of the system at the time of collision. Particles within each cell undergo collisions that change their velocities. The postcollision velocity of particle $i$ in a cell $\xi$ is given by ${\bf v}_i'={\bf V}_{\xi} + \hat{\omega}_{\xi} ({\bf v}_i - {\bf V}_{\xi})$, where ${\bf V}_{\xi}$ is the
center of mass velocity of particles in the cell and $\hat{\omega}_{\xi}$ is the rotation operator of the cell $\xi$. The irreversible chemical reaction $A +C \to B +C$ will eventually consume all of the $A$ fuel. In order to maintain a steady state a bulk phase reaction $B \to A$ is introduced. MPC dynamics has been generalized to incorporate such bulk phase reactions~\cite{rohlf_et_al_rmpcd_2008}. Specifically, at each MPC time step $\tau$ the reaction $B \to A$ is taken to occur probabilistically in a way that depends on the occupancy of the collision cell. The full dynamics of the reacting solvent interacting with the catalytic sphere is described by combining molecular dynamics (MD) for the sphere with reactive MPC dynamics for the solvent~\cite{malevanets_kapral_mpcd_2000}. We also observe that rotational Brownian motion of the sphere is absent since the particles are structureless and central interaction potentials are employed. The hybrid MD-MPC dynamics includes fluctuations, conserves mass, momentum and energy, and accounts for coupling between the $C$ sphere motion and fluid flows.

Simulations were carried out in a cubic box with periodic boundary conditions containing the active $C$ sphere and solvent particles. For multiparticle collisions, carried out at intervals $\tau=0.5 \tsim$, the box was partitioned into $N_c$ cubic cells of linear size $a$. Other parameters are: the average number of solvent particles per cell, $n_0=10$; temperature, $k_B T = 1/3$; solvent mass, $m_A=m_B=m=1$; bulk reaction rate constant, $k_2 = 10^{-3}$. Units of length $a$, mass $m$, energy $\esim$ and time $\sqrt{m a^2/\esim} \equiv \tsim$ are used in the simulations. The $C$ sphere mass is $M=\frac{4}{3}\pi n_0 \sigma^3$. Averages were obtained from 40 realizations of the dynamics.

\begin{figure}[htbp]
  \centering
  \includegraphics[width=\linewidth]{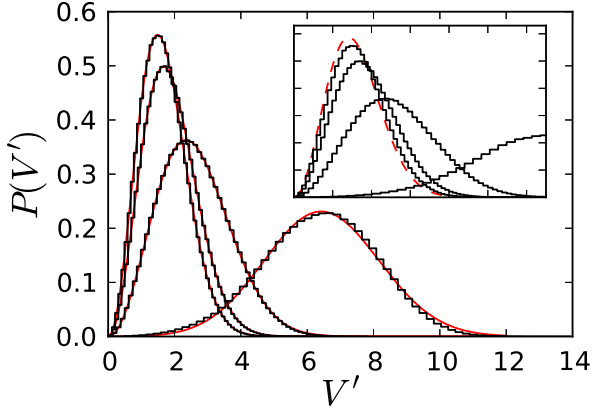}
  \caption{Speed distributions $P(V')$ for spheres of various sizes. The binned
    data is from reactive simulations with $\eca=0.1$, $\ecb=1$ and $N_c=64^3$
    and the red lines are fits using Eq.~(\ref{eq:speed-dist}). Other parameters
    are given in the text, with $\esim=\ecb$. Curves from left to right:
    $\sigma=3, 5, 7$ and $9$. Inset: The binned data is from reactive
    simulations and the red dashed line is the equilibrium Maxwellian
    distribution. For $\sigma=3$, the simulation data is close to the
    Maxwellian.}
  \label{fig:P_of_v}
\end{figure}

By fixing the interaction energies so that $\eca \ll \ecb$ and changing the sphere size $\sigma$, the possibility of a bifurcation leading to self-propulsion can be explored. Figure~\ref{fig:P_of_v} presents a comparison of the simulated speed distributions, $P(V')$ for various values of $\sigma$ (or mass $M$). The speed is scaled, $V'=V/\sqrt{k_BT/M}$, so that the equilibrium speed distributions are the same for all $\sigma$.
For $\sigma=3$ the equilibrium and simulated distributions nearly coincide (see inset) but as $\sigma$ increases large-magnitude deviations are observed. At the largest value of
$\sigma=9$, the peak of the nonequilibrium speed distribution differs considerably from that of its equilibrium counterpart. These results are consistent with a velocity probability distribution that is a Gaussian  $P({\bf V}') \sim \exp{(-|{\bf V}'-{\bf V}_C^\prime|^2/2 w^2)}$. The direction of the velocity will not persist, since strong enough fluctuations will be able to destroy the local concentration inhomogeneity. A new fluctuation will cause the instability to re-occur with the local concentration inhomogeneity near a different portion of the sphere surface. For this reason the observable that most robustly captures the effect and is easily extracted from the simulations is the speed. The speed distribution derived from the velocity distribution is
\begin{equation}\label{eq:speed-dist}
P(V')=\frac{1}{\sqrt{2 \pi} w} \frac{V'}{V_C'} \left( e^{-(V'-V_C')^2/2 w^2}- e^{-(V'+V_C')^2/2 w^2}\right),
\end{equation}
where both $V_C'$ and $w$ vary with the sphere size. As shown in the figure, the fits of this equation, with parameters $V_C'$ and $w$ given in Table~\ref{tab:speed_msd}, are indistinguishable from the simulation data.
\begin{table}[h]
  \centering
  \caption{\label{tab:speed_msd}
    Speed distribution parameters, and ballistic and diffusive components of the MSD. Parameters are the same as in Fig.~\ref{fig:P_of_v}.
  }
    \begin{tabular}{l | c c c c c c c}
    $\sigma$ & 3 & 4 & 5 & 6 & 7 & 8 & 9 \\
    \hline
    $V_C'$           & 0.04 & 0.4 & 0.6 & 0.9 & 1.3 & 2.6 & 5.9 \\
    $w$              & 1.1 & 1.1 & 1.1 & 1.2 & 1.4 & 1.7 & 1.8 \\
    $\langle V' \rangle$ & 1.7 & 1.8 & 1.9 & 2.1 & 2.6 & 3.7 & 6.5 \\
    $V_B'$           & 1.7 & 1.9 & 1.8 & 2.0 & 2.6 & 3.8 & 6.5 \\
    $\rule{0pt}{1.4em}\frac{D_{C}}{10^{-3}}$
                     & 11.4 & 12.6 & 16.2 & 24.0 & 55.9 & 223.4 & 843.3 \\
    $\rule{0pt}{1.4em}\frac{D_C^N}{10^{-3}}$
                     & 5.64 & 3.28 & 2.95 & 1.80 & 1.41 & 1.35 & 0.93
  \end{tabular}
\end{table}

\begin{figure}[htbp]
  \centering
  \includegraphics[width=\linewidth]{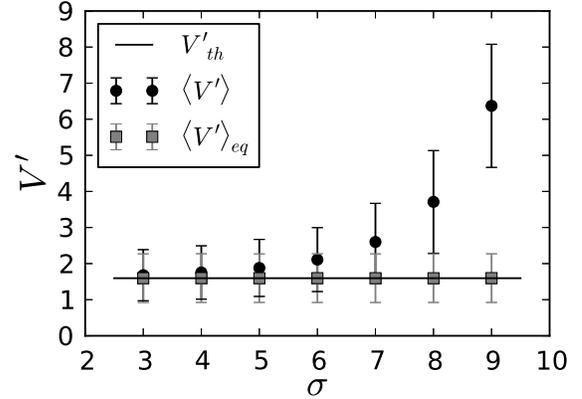}
  \caption{The average velocity $\langle V' \rangle$ of spheres of varying sizes determined from simulation of reactive nonequilibrium and nonreactive equilibrium systems. Parameters are the same as in Fig.~\ref{fig:P_of_v}. The full line is the average thermal velocity, $V_{th}'$, on which all equilibrium simulations fall.}
  \label{fig:velocity}
\end{figure}

In Figure~\ref{fig:velocity} we compare the average speeds $\langle V' \rangle$ for reactive and nonreactive spheres of varying size.
The nonreactive simulations were carried in a solvent consisting only of species $A$ fluid particles. Without reactive events, the system, consisting of the fluid and the sphere, is at equilibrium and one expects the the average speed to be given by $V_{th}'=\sqrt{8/\pi}$, which is indeed seen in the figure.
For the nonequilibrium reactive systems a strong departure of $\langle V' \rangle$ from the thermal value is observed as $\sigma$ increases. The average
speed computed from Eq.~(\ref{eq:speed-dist}) is
\begin{equation}
\langle V' \rangle= \sqrt{\frac{2}{\pi}} w e^{-V_C^{{\prime}2}/2w^2} +  \frac{V_C^{{\prime}2}+ w^2}{V_C'} {\rm erf}{(V_C'/\sqrt{2} w)},
\end{equation}
which varies from the thermal speed, $V_{th}'$, when $\sigma$ is small and $V_C'=0$, to $V_C'$ for large $\sigma$.
These results are consistent with a bifurcation to self-propelled motion for sufficiently large sphere sizes between $\sigma=4$ and 5.

The nature of the instability and an estimate of the sphere size $\sigma$ at which it occurs can be obtained by considering the force on the sphere at coordinate ${\bf R}$. The instantaneous microscopic force on the sphere is ${\bf F}=  \sum_{\alpha \in \{A,B\}} \int d{\bf r} \; \rho_\alpha({\bf r}; {\bf r}^N) (\partial V_{C\alpha}(|{\bf r}-{\bf R}|)/\partial{\bf r})$, where the microscopic density of species $\alpha$ at point ${\bf r}$ is $\rho_\alpha({\bf r}; {\bf r}^N)= \sum_{i=1}^{N_\alpha} \delta({\bf r}-{\bf r}_{i\alpha})$ and ${\bf r}_{i\alpha}$ is the coordinate of particle $i$ of species $\alpha$.  A microscopic boundary layer surrounds the sphere within which the intermolecular forces on the sphere act. Due to momentum conservation and the finite range of the potentials, the sphere plus solvent in the boundary layer is force free, which has been used to write this expression for the force. 

Although a full theoretical description of the instability would have to account for fluctuations and the particle-based nature of the simulations, a  deterministic description that utilizes a reaction-diffusion description of the concentration fields can be used to describe the basic underlying mechanism for the instability in theoretical terms and make rough predictions of when it should occur. Consider the average of the force on the sphere over a nonequilibrium ensemble where the sphere has position ${\bf R}$ and velocity ${\bf V}$ and the force $\langle {\bf F} \rangle$ is given by
\begin{equation}
\langle {\bf F} \rangle= \hat{{\bf V}} \sum_{\alpha \in \{A,B\}} \int d{\bf u}\;
\rho_\alpha({\bf u},{\bf V})(\hat{{\bf V}} \cdot \hat{{\bf u}})\frac{dV_{C\alpha}(u)}{du},
\end{equation}
where ${\bf u}={\bf r}-{\bf R}$, $\hat{{\bf V}}$ is a unit vector in the direction of ${\bf V}$ and $\rho_\alpha({\bf u},{\bf V})=\langle \rho_\alpha({\bf r}; {\bf r}^N) \rangle$ is the nonequilibrium average of the microscopic density of species $\alpha$. Outside the boundary layer with outer radius $R_0$, a continuum description of the solvent is assumed to be appropriate. Consequently, we approximate the density as $\rho_\alpha({\bf u},{\bf V}) =
n_\alpha({\bf u},{\bf V})$ for $u > R_0$, and
$\rho_\alpha({\bf u},{\bf V}) = \exp{(-\beta V_{C\alpha}(u))}n_\alpha(\hat{{\bf
    u}}R_0,{\bf V})$ for $u \le R_0$,  where $n_\alpha({\bf u},{\bf V})$ can
be found from the solution to a reaction-diffusion equation. The nonequilibrium average force then takes the form, $\langle {\bf F} \rangle=\hat{{\bf V}} \frac{2}{\beta} \lambda^2 \int d\hat{{\bf u}} \; n_B(R_0 \hat{{\bf u}},{\bf V}) (\hat{{\bf V}} \cdot \hat{{\bf u}})$, where $\lambda^2=\int_0^{2^{1/6}\sigma} du \; u \left( e^{-\beta V_{CB}(u)} -  e^{-\beta V_{CA}(u)} \right)$ accounts for interactions between the reactive species and the sphere.

The location of the instability can be determined following the analysis in Refs.~\cite{mikhailov:97,mikhailov_calenbuhr_from_cells_to_societies_2006}. For a sphere at position ${\bf R}(t)$ the $B$ density field at a point ${\bf r}$ outside the boundary layer satisfies
\begin{equation}
   \label{eq:source}
   \partial_t n_B(\vr,t) = D \nabla^2 n_B(\vr, t) - k_2 n_B +{\mathcal S}({\bf r},t).
\end{equation}
Near the onset of the instability where the velocity is small the Peclet number Pe~$= VR/D$ will not be very large and advective terms can be neglected. The source term ${\mathcal S}({\bf r},t)=(4\pi R_0^2)^{-1}k_0 n_A({\bf r},t)\delta(|{\bf r}-{\bf R}(t)|-R_0)$,
where $k_0$ is the intrinsic reaction rate coefficient for the
reaction $A \to B$ on the $C$ sphere. The formal solution of
Eq.~(\ref{eq:source}) is
\begin{equation}  \label{eq:green}
   n_B(\ur, t) = \int d\ur' dt' G( \ur-\ur' , t-t' ) {\mathcal S}({\bf r}',t')
\end{equation}
where $G(\ur,t) = (4\pi D t)^{-3/2} \exp({-(\ur^2/4 D t + k_2 t)})$ is the
Green function. In order to obtain an approximate expression for $n_B(\ur, t)$ we replace the source term by its lowest order term in the multipole expansion,
${\mathcal S}({\bf r},t) \approx {\mathcal S}_0 \delta({\bf r}-{\bf
  R}(t))$. Substitution into Eq.~(\ref{eq:green}) and evaluation of the
integrals, along with the assumption that the sphere moves at a constant
velocity, yields an asymmetrical $B$ concentration field,
\begin{equation}
   \label{eq:nb}
   n_B(\ur) = \frac{{\mathcal S}_0}{4\pi D u} e^{-{\bf u}\cdot \tilde{{\bf V}}}
   e^{-\sqrt{\kappa^2 + \tilde{V}^2}u},
\end{equation}
where $\tilde{{\bf V}} = {\bf V}/(2 D)$ and $\kappa=\sqrt{k_2/D}$. The $A$
density is given by $n_A(\ur)=n_0-n_B(\ur)$, assuming total density $n_0$ variations
are negligible. Using these results, for small sphere velocities where our approximations are valid, the nonequilibrium average force on the sphere to order $V^3$ is $\beta \langle \vF \rangle  = {\mathcal A} (1-{\mathcal B} V^2) {\bf V}$, where ${\mathcal A} = \frac{4 \pi}{3}  \frac{R_0^2}{D^2}|\lambda^2|r_f$,
with $r_f=(4 \pi R_0^2)^{-1}k_0k_D n_0/(k_0+k_D(1+\kappa R_0))$ the reaction rate per unit area, $k_D= 4 \pi D R_0$ the Smoluchowski rate coefficient, and ${\mathcal B}= \kappa R_0(1-\kappa R_0/5)/(8 k_2D)$.
In writing the expression for the force we used the fact that for our instability condition $\epsilon_{CB}>\epsilon_{CA}$ and $\lambda^2 <0$, and have taken the source strength to be ${\mathcal S}_0 \approx k_0k_D n_0 e^{\kappa R_0}/(k_0+k_D(1+\kappa R_0))$, its value for a stationary sphere.

The instability threshold is determined from the condition where $\langle {\bf F} \rangle$ exceeds the frictional force leading to a growth of the velocity instead of its decay. Letting $\langle {\bf F} \rangle/\zeta= {\mathcal C} {\bf V}$, where $\zeta$ is the friction coefficient and
\begin{equation}\label{eq:C}
{\mathcal C} = \frac{4 \pi}{3}  \frac{k_BT}{\zeta} \frac{R_0^2}{D^2}|\lambda^2|r_f ,
\end{equation}
the instability condition is ${\mathcal C}=1$, with instability for ${\mathcal C} > 1$. For our fixed potential parameters the instability occurs at $\sigma \approx 4.7 $, which is consistent with the threshold range estimated from the simulations in Fig.~\ref{fig:velocity}. The friction coefficient was found from the decay of the sphere velocity correlation function for nonreactive systems.

Nonlinear terms in the expression for $\langle {\bf F} \rangle$ will lead to
saturation of the instability and the final self-propelled velocity of the
sphere. Beyond but close to the instability threshold the velocity is
given by $V_C^2=({\mathcal C} -1)/({\mathcal C} {\mathcal B})$. The sphere
velocities for several $\sigma$ values, including values far beyond the
instability threshold where the analytical estimates break down, are given in
Table~\ref{tab:speed_msd}. For example, for $\sigma=5$ the theoretical estimate
gives $V_C' \approx 1.9$, which is comparable to but higher than $V_C'=0.6$ and
is close to $\langle V' \rangle$ in the Table for this $\sigma$.

The velocity of the sphere is not constant as assumed in these theoretical estimates and it experiences
fluctuations in its norm (cf. Fig.~\ref{fig:P_of_v}) and direction as a result of local concentration fluctuations. Consequently, diffusive motion will be observed on long time scales. The mean square displacement (MSD), $\Delta L^2(t)$, was used to characterize
the short-time ballistic motion and long-time diffusive behavior of the sphere
under reactive nonequilibrium and nonreactive equilibrium conditions. The MSD of
the nonreactive simulations is described by $\Delta L^2(t)=
6D_C^N(t-\tau_V(1-e^{-t/\tau_V}))$, where the velocity relaxation time
$\tau_V=M/\zeta$. The crossover from short-time inertial motion (present in our MD-MPC dynamics), $\Delta L^2(t) \sim
3(k_BT/M)t^2$, for times $t \ll \tau_V$ to diffusive behavior, $\Delta L^2(t)
\sim 6D_C^N t$, for times $t \gg \tau_V$, where $D_C^N=k_BT/\zeta$, occurs at a
crossover time $t_c \approx 2 \tau_V$.

\begin{figure}[htbp]
  \centering
  \includegraphics[width=\linewidth]{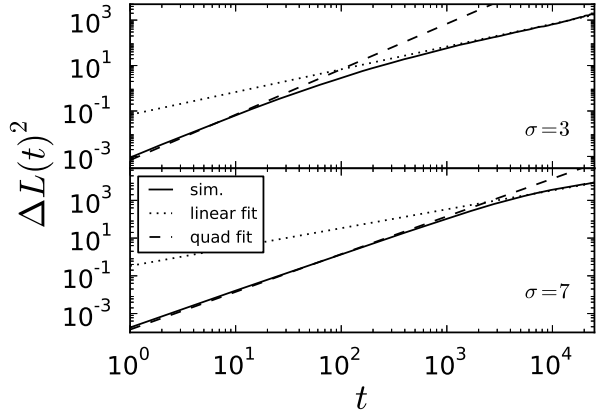}
  \caption{Log-log plot of $\Delta L^2(t)$ versus $t$ for active spheres with $\sigma=3$ and 7 from an average over 40 realizations of the dynamics. The straight dashed and dotted lines indicate the ballistic and diffusive regimes, respectively. Their intersection yields an estimate of the crossover time $t_c$.}
  \label{fig:msd}
\end{figure}
The MSD for chemically active spheres also displays ballistic, $\Delta L^2(t) \sim V_B^2 t^2$, and diffusive, $\Delta L^2(t) \sim 6D_C t$, components (see Fig.~\ref{fig:msd}) but the values of these components differ from those of the nonreactive simulations, and the crossover time is up to two orders of magnitude larger than $\tau_V$. Table~\ref{tab:speed_msd} lists the values of the diffusion coefficients $D_C$ and $D_C^N$ for reactive and nonreactive systems, respectively, for various values of $\sigma$, obtained from fits of the MSD. The Table also gives the value of $V_B$ in the ballistic contribution. While $D_C^N$ decreases with $\sigma$, $D_C$ shows a very strong increase with $\sigma$ for large $\sigma$, consistent with self-propelled motion. From the Table we see that the ballistic speed $V_B$ tends to the inertial value $V_B \sim \sqrt{3 k_BT/M}$ for small $\sigma$ and to $V_B \sim V_C$ for large $\sigma$, again consistent with self-propelled motion. The Reynolds numbers corresponding to the chemically active simulations are less than unity indicating viscous rather than inertial effects dominate the dynamics.

The approximate expression for ${\mathcal C}$ in Eq.~(\ref{eq:C}) can be used to estimate when an instability leading to self propulsion will be likely to occur in physical systems. Since such estimates are system-specific we discuss general system characteristics that favor the instability. Taking $\zeta$ to be given by its Stokes value, $\zeta=6\pi \eta R_0$, the instability condition ${\mathcal C}=1$ reads $\frac{2}{9} \frac{k_BT}{\eta}\frac{R_0}{D^2}|\lambda^2|r_f=1$. For typical values $k_BT \sim 4 \times 10^{-21}$ kg m$^2$/s, $\eta \sim, 10^{-3}$ kg/ms and $D \sim 10^{-9}$ m$^2$/s, the factor $\frac{2}{9} \frac{k_BT}{\eta}\frac{1}{D^2} \sim 1$ s/m. Consequently, we require $R_0|\lambda^2|r_f>1$~m/s for instability. The microscopic length $|\lambda|$ depends on the intermolecular potential and will often have values ranging from Angstroms to nanometers. The reaction rate, which can be controlled by varying the concentration $n_0$ and other factors, is an important quantity to consider for the instability. When reaction with the sphere is the rate controlling step, $r_f= k_0 n_0/(4\pi R_0^2)$ and, since $k_0 \sim R_0^2$, $r_f$ is independent of $R_0$. In contrast, for diffusion control $r_f= k_D n_0/(4 \pi R_0^2)=D n_0/R_0$. Systems with such characteristics where the instability mechanism operates could be investigated experimentally.  For instance, for reaction control with $r_f \sim 10^{23}$ molecules/m$^2$s, instability will occur for $R_0 \sim 10 -100$ $\mu$m.

The velocity increases from zero above the bifurcation point and, close to the
bifurcation, the formula $V_C^2=({\mathcal C}-1)/({\mathcal C}{\mathcal B})$ can
be used to estimate its value. The parameter ${\mathcal C}$ controls the
distance from the bifurcation while the parameter ${\mathcal B}$ depends on the
radius $R_0$, the diffusion coefficient $D$ and the inverse length $\kappa$ that
gauges how bulk reaction in the environment destroys product molecules to set up
a steady state. Typically $\kappa R_0 <1$. Using the parameters for the reaction
controlled case described above, and taking ${\mathcal C} \approx 1.1$, not too
far above the bifurcation point so that the formula retains its validity, we find
$V_C \approx 3-30 \mu m/s$, values which are similar to those for (smaller)
self-propelled particles with asymmetric catalytic activity. Of course, accurate
estimates of the velocity will depend on the specific details of the particular
system under study.

Recently the diffusion coefficients of catalytically active enzymes were observed to be substantially larger than their inactive counterparts~\cite{muddana_et_al_enzyme_jacs_2010,sengupta:13}. In particular for catalase, an enzyme with one of the fastest turnover rates, the diffusion coefficient increased by 45\%
in $H_2O_2$ solution where active catalysis takes place. Assuming that our theoretical instability condition holds for catalase and
using parameters appropriate for this enzyme, we estimate that ${\mathcal C}$ is well below the instability threshold. However,
even below the instability threshold reactive fluctuations can lead to enhanced diffusion~\cite{golestanian_prl_2009}. For example, for our reactive system with $\sigma=3$, $\eca=1$ and $\ecb=4$, which is below the instability threshold, the diffusion coefficient is found to be $D_C= 7.0 \times 10^{-3}$, while the diffusion coefficient for a nonreactive system with the same parameters is $D_C^N= 5.5 \times 10^{-3}$. There is a 27\% increase for the reactive system. These results are consistent with the experimental observations and interpretations of this effect. In addition to these results on enzymatic systems, recently experiments have shown that simple catalytic Pt spherical and composite particles exhibit enhanced diffusion and ballistic motion when $H_2O_2$ is present in solution~\cite{yamamoto:13}. These results are also in accord with our simulations on reactive dynamics below the instability threshold.

 Our results provide a molecular-based demonstration of self-propulsion through symmetry breaking that incorporates the effects of reactive concentration fluctuations and hydrodynamic flows. Further, they demonstrate the existence of enhanced diffusion, even below the instability threshold, and suggest mechanisms for enhanced diffusion in active enzymatic systems.

\acknowledgments
Research of RK was supported by NSERC and a Humboldt Research Award. Computations were performed on the GPC supercomputer at the SciNet HPC
Consortium~\cite{scinet-short}.  PdB would like to acknowledge interesting discussions with J.-P. Boon.

\bibliographystyle{eplbib}
\bibliography{ballistic}

\end{document}